\documentclass[prl,aps,showpacs,amsmath,amssymb,twocolumn]{revtex4-1}

\usepackage{graphicx}

\begin{document}

\preprint{PREPRINT}

\title{The effect of confinement on the interaction between two like-charged rods}

\author{G. Odriozola} 

\author{F. Jim\'{e}nez-\'{A}ngeles} 

\author{M. Lozada-Cassou} 

\affiliation{Programa de Ingenier\'{\i}a Molecular, Instituto
Mexicano del Petr\'{o}leo, L\'{a}zaro C\'{a}rdenas 152, 07730
M\'{e}xico, D. F., M\'{e}xico}

\date{\today}
\begin{abstract}
Monte Carlo simulations were employed to study two charged rods confined between two unlike charged plates, all immersed in a model electrolyte. Recently, it was shown that two rods immersed in a multivalent counterion solution may show attraction (PRL \textbf{78}, 2477 (1997)). Here we show for a monovalent electrolyte that rod-rod attraction and repulsion can switch sign depending on confinement and ionic size. We also propose a simple self-assembling mechanism which may be helpful to understand the DNA-lipid bilayers complexation.
\end{abstract}

\pacs{87.15.Kg,61.20.Gy,61.20.-p}

\maketitle

Confinement effects appear in a wide range of length scales. They are observed from very large scales \cite{Namouni}, up to subatomic scales \cite{Rauch}, passing though molecular levels \cite{Israelachvili,Levinger}. At molecular levels, confinement induces ordering, produces density fluctuations, and provokes dielectric changes on liquids, and even enhances miscibility in polymer blends \cite{Israelachvili,Levinger,Zhu}. In colloidal science, it is known to produce attractions that have yet to be explained \cite{Han}, and in biological setups, the presence of plate-like, parallel, charged lipid bilayers was shown to induce attraction between like-charged DNA molecules \cite{Wagner00,Radler97,Wu04M,Safinya01,Liang04}.

In soft-matter and at nanoscale levels, confinement affects many body electrostatic forces, short range correlations, and thus, macroparticles double layer interactions. Additionally, other associated phenomena such as counterion release appear, as a function of the distance from the macroparticles to the approaching, unlike charged, confining surfaces. Counterion release is understood as the expulsion of counterions from the space between the macroparticles and the approaching surfaces \cite{Wagner00,Radler97,Wu04M,Safinya01,Liang04,Sens00}, and it is claimed to be the driving force for higher-order self assembling \cite{Safinya01,Radler97,Liang04}. This conclusion was built from DNA lipid bilayers complexation experiments, assuming that the cationic lipid bilayer fully neutralize the DNA charge. As we will show, this condition is not fulfilled. For the lamellar phase, this system may be modeled by a couple of negatively charged rods (DNA), sandwiched by parallel positive charged plates (bilayers). Nevertheless, aside from important contributions dealing with parallel rigid polyelectrolytes \cite{Gronbech97,Lowen94,Lee04,Ha,Allahyarov,Felipe4}, there are no theoretical works on the effects that confinement may produce on their interaction. Thus, we perform Monte Carlo (MC) simulations of this system for elucidating these effects and the associated phenomena.

\begin{figure}
\resizebox{0.35\textwidth}{!}{\includegraphics{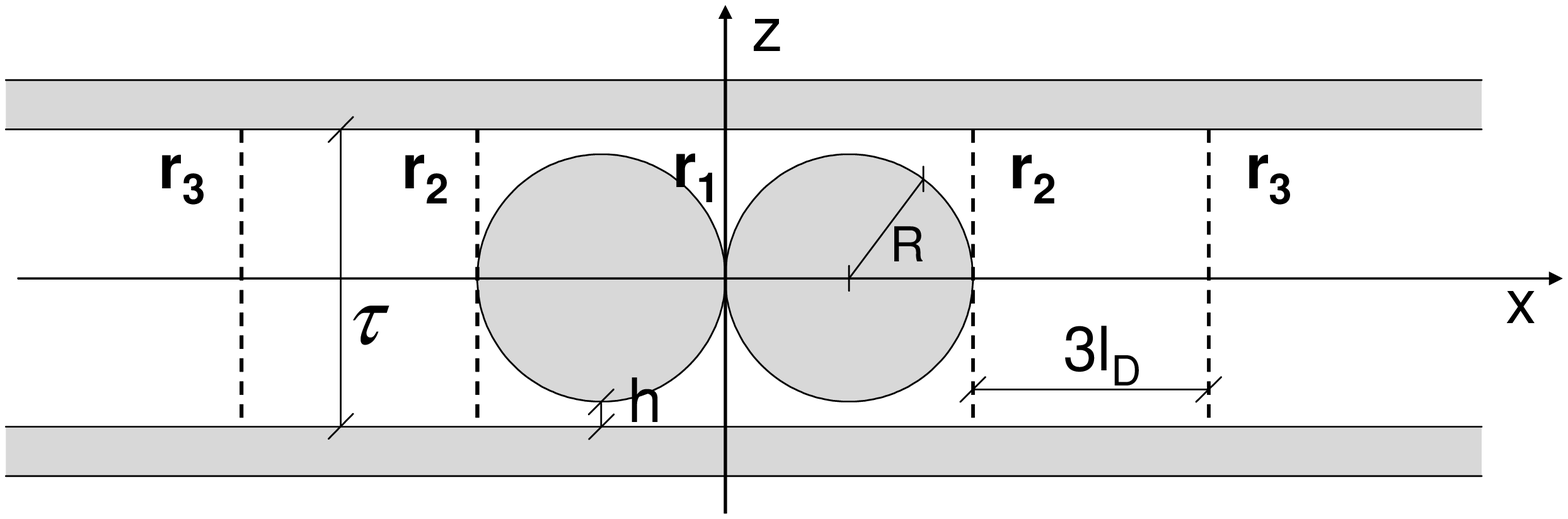}}
\caption{\label{scheme} Scheme of the system showing regions $r_1$, $r_2$, and $r_3$. }
\end{figure}

$NVT$ MC simulations were performed for two infinitely large, hard, parallel, negative rods confined between two hard, parallel, positive plates of finite thickness, which are immersed in a 1-1 restricted primitive model (RPM) electrolyte, unless otherwise specified (see Fig.~\ref{scheme}). RPM consists of hard spheres with a centered point charge, such that their electrostatic interaction is $U_{E}(r_{ij})$$=$$\frac{\ell _{B}z_i z_j}{\beta r_{ij}}$, being $\beta$$=$$1/k_BT$, $k_{B}$ the Boltzmann constant, $T$$=$$298$K the absolute temperature, $z_i$ and $z_j$ the valences of sites $i$ and $j$, $\ell _{B}$$=$$\frac{\beta e^2}{\epsilon}$$=$$7.14$$\hbox{\AA}$ the Bjerrum length, $\epsilon$$=$$78.5$ the dielectric constant, and $r_{ij}$ the interparticle distance. For simplicity, plates, rods, and solution have the same dielectric constant. The simulation box has sides lengths $L_x$$=$$L_z$$=$ 200 $\hbox{\AA}$, and $L_y$$=$ 125 $\hbox{\AA}$. The origin of coordinates is set at the box center. Plates are located parallel and symmetric to the $z$$=$$0$-plane, with a surface-surface separation distance $\tau$. Rods axes are placed parallel to the $y$ axis at $x$$=$$\pm (R+t/2),z$$=$$0$, being $R$ the rods radii and $t$ the rods surface distance. We assigned to plates a charge density of $\sigma_0$$=$ $0.229$$C/m^2$ ($+e$ per 70 $\hbox{\AA}^{-2}$) for each surface, in correspondence with a fully charged cationic lipidic membrane \cite{Gelbart00}. We place $+e$-charges over a triangular grid, which is randomly moved in the $x$ and $y$ directions to mimic a continuous charge distribution. Rods have an $-e$-site each 1.7 $\hbox{\AA}$, as DNA molecules \cite{Gelbart00}. Rods have a radius of $R$$=$10.5 $\hbox{\AA}$, consistent with a hydrated DNA molecule. Plates are 5 $\hbox{\AA}$ thick to allow fluid-fluid correlations between the interplate and bulk sides of the plates \cite{marcelo96a}. These correlations are expected in real 40 $\hbox{\AA}$ thick bilayers, due to their low dielectric constant. We fixed the concentration of the 1-1 electrolyte to $\rho_s$$=$0.1 M, and assigned its diameter to $a$$=$4.25 $\hbox{\AA}$, unless otherwise indicated. Additional anions are added to make the system electroneutral. Periodical boundary conditions are set for all directions. The Ewald summation formalism is employed to deal with Coulomb interactions \cite{Odriozola_lan,Alejandre94}. To access all phase space volume, {\it i.~e.}, to allow for ion interchange between the confined and unconfined regions, movements having a large maximum displacement are done. Electrostatic contributions to the forces acting on the macroparticles (rods and plates) are obtained by $\mathbf{F}_{el}$$=$$\langle \sum_i \sum_j -\nabla U_{E}(r_{ij}) \rangle$, where $i$ runs over the sites of the reference macroparticle and $j$ runs over all other sites. The contact contribution is obtained by integrating the ions contact density,$\rho(s$$=$cte$)$, $\mathbf{F}_c$$=$ $-\int_sk_{B}T\rho(s$$=$cte$)\mathbf{n}ds$, where $s$ refers to the macroparticle surface and $\mathbf{n}$ is a unit normal vector. These two contributions to the force are interdependent.

\begin{figure}
\resizebox{0.44\textwidth}{!}{\includegraphics{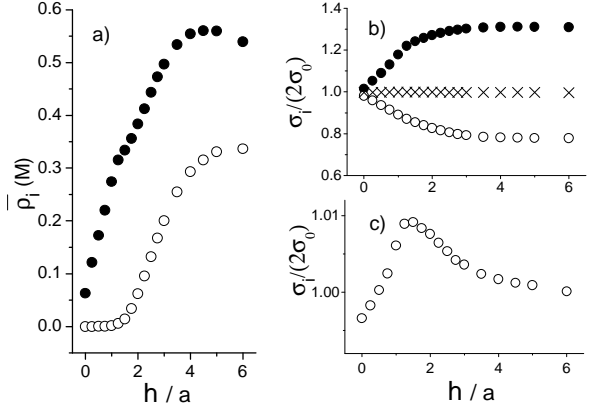}}
\caption{\label{charge-cations} a) Cation, $\circ$, and anion, $\bullet$, densities for region $r_1$. b) Normalized charge densities for regions $r_1$  ($\bullet$), $r_2$ ($\circ$), and ($r_3$) $\times$. c) Normalized charge density of region $\mid \! z \! \mid$ $\leq$ $\tau/2$. }
\end{figure}

\begin{figure}
\resizebox{0.47\textwidth}{!}{\includegraphics{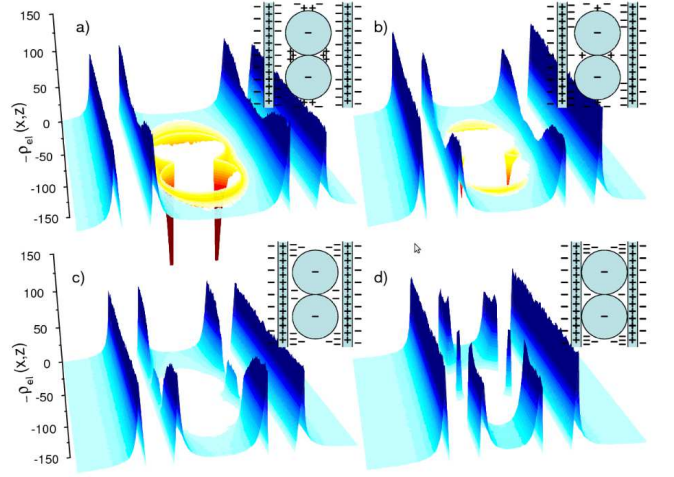}}
\caption{\label{rhos} Negative of $\rho_{el}(x,z)$ for $h/a=$ 2.5 (a), 1.5 (b), 1.0 (c), and 0 (d). Blue indicates a larger concentration of anions than cations and red the opposite. Darker tones indicate larger absolute values of $\rho_{el}(x,z)$. The inset of each figure schematizes the charge distribution. }
\end{figure}

First simulation experiment consists of studying the system as a function of $h$$\equiv$$(\tau-2R)/2$, for $t$$=$$0$, being $h$ the rod-plate surfaces distance. To study the ion release mechanism we focused on the following interplate ($\mid \! z \! \mid$ $<$ $\tau/2$) regions: ($r_1$) $\mid \! x \! \mid$ $<$ $2R$, ($r_2$) $2R$ $\leq$ $\mid \! x \! \mid$ $<$ $2R$$+$$3l_D$, and ($r_3$) $2R$$+$$3l_D$ $\leq$ $\mid \!x \!\mid$ (see Fig.~\ref{scheme}), being $l_D$$=$$\{\epsilon /(2 \beta e^2 \rho_s)\}^{1/2}$$=$$9.6$$\hbox{\AA}$ the Debye length. The average density of cations and anions, $\bar{\rho}_i$$=$ $\int_{r_1}\rho_i(\mathbf{r})d\mathbf{r}$, are presented in Fig.~\ref{charge-cations} a). Figure \ref{charge-cations} b) shows the normalized charge density, $\sigma_{r_j}/(2\sigma_0$), for region $r_j$, where $\sigma_{r_j}$$=$$\frac{1}{A_j}\int_{r_j}\rho_{el}(\mathbf{r})d\mathbf{r}$, $A_j$ is the area of the $j$ region projected on the plate, and $\rho_{el}(\mathbf{r})$ is the volume charge density including anions, cations, and rods. Figure \ref{charge-cations} c) shows the normalized charge density of the interlayer region, $\sigma_{in}/(2\sigma_0)$, being $\sigma_{in}=\frac{1}{A_t}\sum_{j}A_{j}\sigma_{r_j}$ and $A_t$$=$$\sum_{j}A_{j}$. In Fig.~\ref{charge-cations} b) is seen that in $r_3$ the induced charge practically compensates that on the plates for all $h$. This is something expected since this region is relatively far from the rods, and  plates are relatively far from each other: $\tau$$\geq$$21$$\hbox{\AA}$$\approx$$5a$. On the other hand, local electroneutrality is violated for the other two regions, {\it i.~e.}, the charge of the confined fluid does not compensate the corresponding charge of the plates \cite{marcelo96b}. It is seen that the charge of region $r_1$ always overcompensates it, whereas region $r_2$ always undercompensates it. Both effects, in turn, induce charge polarization in the fluid outside the plates. However, the overall charge of region $r_1+r_2$ does not compensate the charge of plates, but produce violations of charge electroneutrality. This yields the $\sigma_{in}/(2\sigma_0)$ plot shown in Fig.~\ref{charge-cations} c). This effect was predicted by integral equation studies for confined electrolytes. Furthermore, it was also shown that a correlation exists between the inside and outside of the plates \cite{marcelo96a}. Here, we found both effects to be in agreement with those results (see also Fig.~\ref{rhos}). There is undercompensation of the plate's charge for $h$$<$$0.5a$ and overcompensation for $h$$>$$0.5a$ (Fig.~\ref{charge-cations} c)). Electroneutrality is yield for large $h$ values. It is also seen from Fig.~\ref{charge-cations} a) that ion release starts at $h \approx 4a$, as plates approach each other. However, cation release ends at $h$$=$$1.5a$ since region $r_1$ runs out of them. On the contrary, anion release is strongly produced for $h$$<$$1.5a$. Hence, both facts, running out of cations and strongly releasing anions for $h$$<$$1.5a$ produce the decrease of negative charge at region $r_1$ (solid symbols of Fig.~\ref{charge-cations} b)), which in turn manifest as the maximum of $\sigma_{in}/(2\sigma_{0})$ in Fig.~\ref{charge-cations} c). Electroneutrality is violated at the expenses of keeping a constant chemical potential \cite{marcelo96a,marcelo96b}. As we discuss below, $\sigma_{in}$ determines the electrical contributions to the force on plates.

The charge distribution profile around rods and plates is defined by $\rho_{el}(x,z)$$=$$z_{+}g_{+}(x,z;\tau)+z_{-}g_{-}(x,z;\tau)$, being $g_{+}(x,z;\tau)$ ($g_{-}(x,z;\tau)$) the cation (anion) distribution profile, given $\tau$; $\rho_{el}(x,z)$ is shown in Fig.~\ref{rhos} for different values of $h$. For large $h$, the plates double layers are uncorrelated from those of the rods and from those outside the plates (not shown). For smaller $h$, the double layers start to interact, becoming strongly correlated. This is already seen in Fig.~\ref{rhos} a), where the positive charge around the rods distorts due to the presence of the plates double layers and viceversa. Consequently, depletion regions appear between rods and plates. As pointed out above, from a mechanistic point of view, these depletion regions are linked to partial release of cations and anions from the interplate region towards the bulk, thus, due to the plates and rods double layers interactions, {\it i.~e.}, the particles charge and size correlations. However, from a fundamental point of view, the depletion regions are due to the need of the system to increase particles accessible volume and, hence, the entropy, and ultimately, due to the conservation of the chemical potential. Figure \ref{rhos} b) is built for $h$$\approx$$1.5a$, where cations were practically expelled out from the interplate region. There, small red peaks recall us the presence of positive ions. Additionally, the anionic depletion regions were largely increased. For smaller $h$, no more cationic peaks are observed (Fig.~\ref{rhos} c) and Fig.~\ref{rhos} d)), and the anionic depletion regions enlarge. This is just a consequence of excluded volume, since no ions can enter the region in between rods and plates. In addition, the regions at $x$$=$0,$z$$=$$\pm\sqrt{a(R+a/4)}$, where cations used to place, are now populated by anions. The insets in Fig.~\ref{rhos} schematize the ionic distribution for these configurations.

\begin{figure}
\resizebox{0.47\textwidth}{!}{\includegraphics{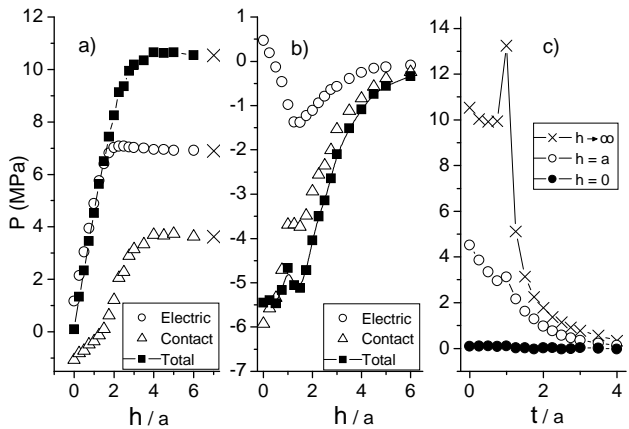}}
\caption{\label{forces-pressures} a) Forces on rods per unit of rod area (pressure) as a function of $h$ for rod-rod contact. Crosses were obtained when no plates were included in the simulation box. b) Forces on plates per unit of plate area (pressure) as a function of $h$. c) Pressure on rods as a function
of $t$ for different $h$. }
\end{figure}

Effective rod-rod forces per unit area (pressures) are shown in Fig.~\ref{forces-pressures}. Note that system symmetry makes the rod-rod force to have only components on $x$. This component is shown in Fig.~\ref{forces-pressures} a), and a positive sign means repulsion. Similarly, just the plate-plate $z$ component of the force, per unit area, is shown in Fig.~\ref{forces-pressures} b). Figure \ref{forces-pressures} a) shows that the influence of plates on rods force is non negligible only for $h$$<$$4a$. For larger separations, the obtained force is equal to that obtained when no plates are included in the system (crosses). This is a consequence of the fact that plates and rods double layers are uncorrelated, as explained above. This force is composed by a large electric repulsive contribution plus a positive contact contribution, which arises as a consequence of the large adsorption of counterions at $x$$=$$0,$$z$$=$$\pm\sqrt{a(R+a/4)}$ \cite{Felipe4}. The contact contribution starts decreasing for $h$$\approx$$4a$ and reaches zero for $h$$\approx$$1.5a$. This decreasing contact contribution is a consequence of the hight decrease of the cations peaks at $x$$=$0,$z$$=\pm\sqrt{a(R+a/4)}$. The attractive contribution of this force for $h$$<$$a$ is due to the anion peaks that grow at the outer sides of the rods, which overcompensate the contribution of the anion peaks at $x$$=$0,$z$$=\pm\sqrt{a(R+a/4)}$. This is seen in Fig.~\ref{rhos} c) and \ref{rhos} d). Finally, the decrease of the electric contribution is due to the increase of the anionic concentration in $r_2$ and $r_3$, which partially compensates the rod-rod non-screened electric force. Both contributions produce an ever decreasing total force, which yields zero for rods-plates contact, $h$$=$$0$.

Figure \ref{forces-pressures} b) shows negative values of the total plate-plate force for all distances, indicating that plates attract each other. For large values of $h$ both contributions to the total force tend to vanish due to the independence of the double layers. As $h$ is decreased the contributions become negative. The electric contribution, however, shows an absolute minimum at $h$$\approx$$1.5a$. Thus, for $h$$<$$1.5a$ it increases yielding a positive value for $h$$=$0. These features recall us the $h$ dependence of $\sigma_{in}/(2\sigma_0)$ (see Fig.~\ref{charge-cations} c)). In fact, both curves are practically equally shaped, indicating that the electrical component of plate pressure is given by $P_{el}$$\sim$$(\sigma^2_{out}-\sigma^2_{in})$, where $\sigma_{out}$ is the induced charge density for the unconfined fluid \cite{marcelo96a}. Hence, counterion release, which is related to the charge density, in turn is linked to the electrical component of the pressure on plates. On the other hand, depletion regions are the cause for the attractive contact contribution to the plates forces. As was shown in Fig.~\ref{rhos}, depletion regions enlarge as $h$ decreases, explaining the general increase of the contact attraction. It should be noted that it is the contact force the main contribution to the net force on plates. In fact, for rod-plate contact, this is the component that yields attraction, in contrast to the electric contribution. Finally, Fig.~\ref{forces-pressures} c) shows the forces on rods per unit area as a function of the rods surface distance, $t$, for different degrees of confinement. It is clearly seen how confinement affects the effective force, producing practically null values for all $t$ and $h$$=$0. In fact, the work per charged site to approach the rods from $t$$=$$4a$ to $t$$=$$0$ is $2.13$$k_BT/site$ for unconfined conditions, $0.78$$k_BT/site$ for $h$$=$$a$, and $0.02$$k_BT/site$ for $h$$=$$0$.

\begin{figure}
\resizebox{0.26\textwidth}{!}{\includegraphics{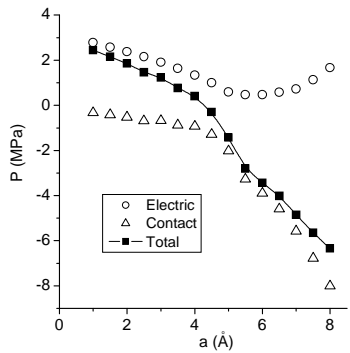}}
\caption{\label{force-ionsize} Force acting on rods as a function of the ionic size.}
\end{figure}

For the parameters employed, experiments show the spontaneous self assembling of bilayers (plates) and polyelectrolytes (rods) into a compact structure \cite{Safinya01}. Our results clearly indicate that plates squeeze the rods.  On the other hand, although our model shows that the necessary energy to pack the rods is practically zero for highly confined conditions, it does not predict attraction. Nevertheless, there is a very important difference between real experiments and our model; that is the employed rod concentration. Experimentalists need a sufficiently large polyelectrolyte concentration to obtain the compact structure: As more polyelectrolyte enters the midplane, the accessible volume fraction decreases, enhancing short range correlations, which in turn produce higher adsorption at surfaces \cite{Felipe1}. This effect may produce a net attractive forces on rods. In order to show this, we did two final computer experiments for $h$$=$$t$$=$$0$. A): we studied the influence of the ion size, $a$. B): we add a macroion species, with valence $-5$ and diameter $13.4$ $\hbox{\AA}$, at the isoelectric point concentration (when the membranes charge equals the macroions charge) \cite{Safinya01}. Here, we also include all plates counterions (valence $-1$, diameter $4.25$ $\hbox{\AA}$) and the necessary cations (valence $1$, diameter $4.25$ $\hbox{\AA}$) to keep the system electroneutral. We wish to mimic a DNA-lipid complex counting on a finite DNA concentration, {\it i.~e.}, we expect the macroions to behave as monomers of the DNA molecule. Hence, the macroions size-charge relation was chosen in such a way that the electric field at their surface equals that at the rods surface.

Results showing the influence of the ion size (A) are in Fig.~\ref{force-ionsize}. It is observed that the total force becomes attractive for $a$$>$$4.25$$\hbox{\AA}$. More importantly, it is clear the influence of the ionic size on the rod-rod effective interaction. That is, the larger the little ions, the lower the accessible volume and so, the system entropy is forcedly reduced. Consequently, the system reacts pushing more ions against the plates and rods surfaces, to gain accessible volume and hence, entropy. From a mechanistic point of view, this translates into a larger contact attractive force. This is seen in Fig.~\ref{force-ionsize}, where the contact contribution clearly rules the total force for a large enough ionic size. In the second experiment (B), we observe (not shown) the macroions to replace the plates counterions and to crowd inside the interplate region and around the rods. Thus, the rods induce macroions lines next to them, which, we argue, would play the role of a finite rods concentration. We obtained an effective rod-rod pressure of $-0.81$MPa, and a plate-plate pressure of $-10.38$MPa. Both are attractive pressures that signature the self-assembling tendency of these systems. Hence, these results suggest that complexation into highly packaged structures should be obtained for a certain concentration of large macroions (polyelectrolytes). This is exactly what experiments show, where small ions leave the confined region in favor of larger ones, to increase the system entropy.

In brief, not only handling valence and concentration of added salts may result in attractive forces between macroparticles \cite{Gronbech97,Lee04,Allahyarov,Ha,Felipe4} but also confinement may produce this effect. It was shown how confinement of rods by opositely charged plates reduces their inherent repulsive effective interaction forces as plates approach each other. This is a consequence of the need of the system to lower the free energy by increasing entropy. Nevertheless, the process is linked to several interesting phenomena such as counterion release, local violations of electroneutrality, and to the appearence of ionic depletion regions. Here we show that attraction between rods and plates depends on a delicate  balance between charge and excluded volume of all particles in the system.


\end{document}